\documentstyle[prb,twocolumn,eqsecnum,aps,epsf]{revtex}
\begin{document}
\draft
\title{Applying voltage sources to a  Luttinger
liquid with arbitrary transmission}
\author{Reinhold Egger and Hermann Grabert}
\address{Fakult\"at f\"ur Physik, Albert-Ludwigs-Universit\"at, 
Hermann-Herder-Stra{\ss}e 3, D-79104 Freiburg, Germany}
\date{Date: \today}
\maketitle
\begin{abstract}
The Landauer approach to transport in mesoscopic conductors
has been generalized to allow for strong electronic
correlations in a single-channel quantum wire.
We describe in detail how to account for external voltage
sources in adiabatic  contact   with a quantum wire
containing a backscatterer of arbitrary strength.
Assuming that the quantum wire is in the Luttinger  
liquid state, voltage sources lead to 
radiative boundary conditions applied to
the displacement field employed in the 
bosonization scheme.  We present the exact 
solution of the transport problem for arbitrary
backscattering strength at the special 
Coulomb interaction parameter $g=1/2$. 
\end{abstract}

\narrowtext

\section{Introduction}

One-dimensional (1D) materials have received much
attention in the past few years.  The discovery
of novel 1D conductors and the failure to model these by
Fermi liquid theory have
raised many interesting questions.  The generic 
behavior of electrons in a 1D conductor is described
by the Luttinger liquid (LL) model,\cite{old,book} when
one considers externally screened 
short-ranged interactions.
Experimental realizations of the phenomena predicted
by the LL model
could be based on carbon nanotubes,\cite{tans,egger}
quantum wires in semiconductor heterostructures 
operated in the single-channel limit,\cite{tarucha} 
long chain molecules,\cite{voit} or
edge states in a fractional quantum Hall bar.\cite{fqh}

In this paper, we study electrical transport in 1D conductors  and 
incorporate external voltage sources adiabatically
connected to the quantum wire. 
We focus on the simplest case of
a spinless  LL described by the
standard interaction parameter $g\leq 1$, where
$g=1$ is the noninteracting limit and a small 
value of $g$ equals strong correlations.  The extension
to spin-$\frac12$ electrons or to nanotubes is
then straightforward.  Taking into account 
backscattering effects due to impurities, 
the generic behavior at low energy scales
 can be studied by 
considering a pointlike scatterer of arbitrary
strength $\lambda$.  Then $\lambda=0$ 
corresponds to a clean conductor, while 
$\lambda\to \infty$ is the limit of perfect 
reflection.   Again the extension to a more
complicated situation, e.g.,
several impurities, is straightforward
and not further discussed here.

\begin{figure}
\epsfysize=2.8cm
\epsffile{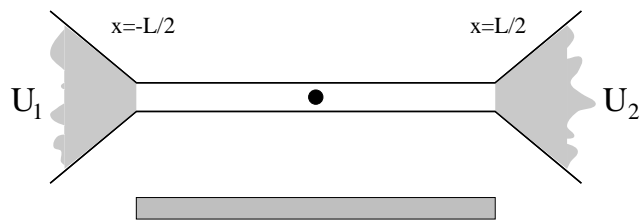}
\vspace{0.3cm}
\caption[]{\label{fig1} 
Gated quantum wire as a model of the Luttinger liquid. 
Reservoirs held at chemical potential $\mu_{1,2}=eU_{1,2}$
are adiabatically connected to the 1D conductor at $x=\mp L/2$.
A backscatterer of strength $\lambda$ (indicated by the 
filled circle) is located at $x=0$. 
The screening gate is responsible for externally screened
interactions within the quantum wire.}
\end{figure}

This underlying geometry is shown in Fig.~\ref{fig1}. 
For clarity, we discuss the case of a gated  single-channel
quantum wire (QW) extending from $-L/2<x<L/2$.
The screening backgate is responsible for short-ranged interactions 
within the QW, and in the single-channel limit
under consideration here, a LL is formed.\cite{glazman,egger2}  
Possibly with minor modifications, however,
the theory applies to all 1D correlated electron
systems.  

At the ends of the
QW, reservoirs are assumed 
to be  adiabatically connected. 
We consider ideal reservoirs as in the
standard Landauer approach
for Fermi liquid conductors.\cite{landauer,datta} 
The reservoirs are held 
at chemical potentials $\mu_{1,2}$, and since 
one has good screening in the (2D or 3D 
Fermi liquid) reservoirs,
the difference $U=(\mu_1-\mu_2)/e$ is the 
applied two-terminal voltage.  For simplicity,
we consider only time-independent voltages in
this paper.  It is shown in Sec.~\ref{sec:bc} that 
the presence of the voltage sources leads
to boundary conditions, which in turn allow
for the application of powerful theoretical techniques, 
e.g., bosonization,\cite{old} refermionization,\cite{book}
or boundary  conformal field theory.\cite{bcft}

Other work has also dealt with
similar questions as the ones addressed here.
(1) There have been attempts to 
describe the effects of a reservoir
by a 1D LL with
$g=1$, both for the clean case \cite{schulz} 
and allowing for impurity backscattering.\cite{furu0}
Albeit such calculations can explain the 
experimentally observed absence
of a conductance renormalization in a perfectly
clean system,\cite{tarucha}
it remains unclear whether this approach can properly
account for ideal reservoirs. Furthermore,
calculations become bulky if one includes
the impurity backscattering.
(2) Other studies have simply assumed a local voltage
drop at the impurity.\cite{kane}   As discussed
in Sec.~\ref{sec:bc}, this assumption is justified
only if the impurity backscattering is effectively very strong. 
(3) In the clean case, Kubo-formula based theories
have been presented\cite{kubo} to explain the perfect
conductance $G=2e^2/h$.
(4) Yet another approach for a clean system models reservoirs
by charges conjugate to the chemical potentials of the
voltage sources.\cite{froh}

We believe that our approach may offer the simplest
and most general answer to the question of
how to incorporate external voltage sources.
It is the natural extension of Landauer's
original approach designed for uncorrelated electrons\cite{landauer} 
to a strongly correlated situation. The theory applies 
for arbitrary interaction strength $g$
and impurity backscattering  $\lambda$.
In particular, as an example of experimental relevance,
we determine the full crossover from perfect conductance 
quantization in a clean wire to the anomalous
power-law conductance suppression in a disturbed wire\cite{kane} 
as the impurity strength is varied.

The structure of this paper is as follows.
In Sec.~\ref{sec:ll} the LL concept is introduced using the gated QW as 
an example.  In Sec.~\ref{sec:bc},
 boundary conditions describing the  
applied voltage sources are presented. Several general consequences 
of the voltage sources are collected in Sec.~\ref{sec:gen}.   
In Sec.~\ref{sec:g12},
the full transport problem for arbitrary 
backscattering strength $\lambda$ is solved
for the special LL parameter
$g=1/2$. Finally, some conclusions
can be found in Sec.~\ref{sec:conc}.
A brief account of some results presented in Secs.~\ref{sec:bc} and 
\ref{sec:gen} has been
given before in Ref.\onlinecite{prl}. 

\section{Luttinger liquid}
\label{sec:ll}

We start by summarizing the
Luttinger liquid concept which allows for
a convenient description of 1D conductors.
Due to Coulomb repulsion, the electrons in a 1D quantum wire
have a tendency to occupy states on a Wigner lattice.
As familiar from lattice excitations, the electronic 
configurations can be described in terms of a 
``displacement field'' $\theta(x)$. 
Remarkably, as far as the low-energy physics
is concerned,  one can always find a one-to-one transmutation
relating the 1D interacting fermion system to an equivalent
bosonic system described by this displacement
field (``bosonization'').
In the bosonization scheme, the right- 
and left-moving component ($p=R/L=\pm)$ of
the electron operator is expressed in terms
of the displacement field,\cite{book}
\begin{equation}   \label{bos}
\psi_p(x)=  (2\pi a)^{-1/2} \exp[-ip k_F x- ip\sqrt{\pi}
\theta(x)-i\sqrt{\pi} \phi(x)] \;,
\end{equation}
where $a\approx 1/k_F$ corresponds to
the lattice spacing of the associated
Wigner lattice and $k_F$ is the Fermi momentum.
The dual field $\phi(x)$ obeys the commutation relation
\begin{equation}\label{algebra}
[ \phi(x),\theta(x') ]_- = -(i/2) \;{\rm sgn}(x-x')\;.
\end{equation}
At low energy scales, any particular dispersion
relation can be linearized around the Fermi energy, and the kinetic energy is
then formally given by a massless Dirac Hamiltonian.
Applying  Eq.~(\ref{bos}) then yields
\begin{equation}
H_0= \frac{\hbar v_F}{2} \int dx 
\left[ (\partial_x \phi)^2 + (\partial_x \theta)^2
\right] \;,
\end{equation}
with  the Fermi velocity $v_F$.  The low-energy
excitations are simply harmonic charge-density wave
oscillations.

In the LL model, one  considers externally screened
short-ranged interactions which, for the purpose
of a low-energy theory, can be represented by
the interaction potential $U_c(x-x')=u_0 \delta(x-x')$.
On length scales larger than the 
screening length imposed by the gate, 
the interaction potential acting in the QW will always
take this form.
The standard LL parameter $g$ is then given by 
\begin{equation}\label{gdef}
g=(1+u_0/\pi \hbar v_F)^{-1/2}\;,
\end{equation}
and the interaction Hamiltonian reads
\begin{equation}\label{hi}
H_I = \frac{u_0}{2} \int dx \,
\rho^2(x)  \;.
\end{equation}
Here $\rho(x)$  is the total density 
relative to the zero-voltage
equilibrium value $\rho=k_F/\pi$.  Including the 
mixed components of the density operator, 
e.g., $\psi_R^\dagger \psi^{}_L$, in Eq.~(\ref{hi}) 
corresponds to retaining the 
electron-electron backscattering.
Following standard arguments,\cite{book}
this is an irrelevant perturbation which can be 
taken into account by a simple renormalization of the
 LL parameters.  The
density $\rho(x)$ entering $H_I$ is then 
only due to the densities $\rho_{R/L}$ of
right- and left-moving fermions, 
$\rho=\rho_R+\rho_L$,
for which Eq.~(\ref{bos})
yields $\rho(x)=\partial_x 
\theta(x)/\sqrt{\pi}$.  Thereby the LL Hamiltonian
emerges,
\begin{equation}
H_{\rm LL} = \frac{\hbar v}{2} \int_{-L/2}^{L/2} dx \left[ 
g(\partial_x\phi)^2
+ g^{-1} (\partial_x \theta)^2 \right] \;.
\end{equation}
For simplicity, we assume full translation invariance 
in the QW such that the sound velocity $v=v_F/g$. 
Remarkably, the low-energy excitations of the
interacting system are still harmonic oscillations, and 
therefore $H_{\rm LL}$ allows for an exact solution.

Next consider a scatterer sitting at, say, $x=0$.
The important backscattering \cite{forfoot} comes 
from the mixed component of the density operator,
$H_{\rm imp}\sim \psi_R^\dagger(0)\psi_L^{}(0) + H.c.$,
which, expressed in terms of the displacement
field, leads to
\begin{equation}
H_{\rm imp} = \lambda \cos[\sqrt{4\pi}\,\theta(0)]\;.
\end{equation}
The energy $\lambda$ is a measure of the impurity
backscattering strength.
The Hamiltonian $H=H_{\rm LL} + H_{\rm imp}$ has
been subject of intense theoretical effort in the past few
years. The purpose of our
paper is to clarify how this
strongly correlated model should incorporate 
applied voltage sources.

For the gated QW in Fig.~\ref{fig1}, the
interaction contribution $H_I$ in Eq.~(\ref{hi})
can be interpreted as the 
charging energy $(e^2/2c)\int dx \, \rho^2(x)$  
of the gate-QW capacitor, where $c=e^2/u_0$ 
is the capacitance per unit length. The electrostatic
potential $\varphi(x)$ in the QW then follows by comparison
with 
\begin{equation}
H_I=(e/2) \int dx\, \rho(x) \varphi(x)
\end{equation}
as
\begin{equation}\label{poisson}
e\varphi(x) = u_0 \rho(x) \;.
\end{equation}
Since we have an effectively short-ranged interaction,
the Poisson equation is replaced by Eq.~(\ref{poisson})
here.  The electrostatic potential 
directly gives the local potential
drop between the QW and the screening backgate.
The noninteracting limit $u_0=0$ ($g=1$)
then implies that the backgate is located within the wire.
The electrostatic potential is thus zero everywhere.
In contrast, if no gate is present, i.e., in the limit of 
unscreened Coulomb interactions,
we can put $g\to 0$ in the long-wavelength
limit.  

Suppose now that the densities
$\rho_R^0$ and $\rho_L^0$ of right- or left-moving
electrons are injected into the QW. This will
charge the gate-QW capacitor and imply a
voltage drop according to Eq.~(\ref{poisson}).
The electrostatic potential (\ref{poisson}) 
shifts the band bottom  by $-e\varphi(x)$.
With the density of states $1/\pi\hbar v_F$ this implies
a shift of the total density by $-e\varphi(x)/\pi \hbar
v_F$.  Therefore the actual density in the QW has to be
self-consistently determined from Eq.~(\ref{poisson})
and the relation
\begin{equation}
\rho = \rho_R + \rho_L = \rho_R^0 + \rho_L^0
-  e\varphi/\pi \hbar v_F \;.
\end{equation}
Using Eq.~(\ref{gdef}) the solution is 
\begin{equation}\label{g2}
\rho(x) =
\rho_R(x) + \rho_L(x) = g^2 [\rho_R^0(x)+\rho_L^0(x)]\;. 
\end{equation}
Since the electrostatic potential is only linked to
the total density via Eq.~(\ref{poisson}), the difference of the $R/L$
moving densities stays invariant,
\begin{equation} \label{rmin}
\rho_R-\rho_L = \rho_R^0 - \rho_L^0 \;.
\end{equation}
This difference determines the current flowing
through the QW,
\begin{equation}  \label{ii}
I= e v_F (\rho_R-\rho_L) \;,
\end{equation}
which can be computed at any point $x$ due to
the continuity equation.    
We note in passing that for the a.c.~case, a displacement current 
has to be added to Eq.~(\ref{ii}).

\section{Voltage sources}
\label{sec:bc}

Next we wish to include the adiabatically connected external  voltage sources
indicated in Fig.~\ref{fig1}.  The left reservoir
held at chemical potential $\mu_1=eU_1$ injects 
the bare density 
\begin{equation}\label{bc1}
\rho_R^0(-L/2)= eU_1 /2\pi\hbar v_F 
\end{equation}
of right-movers into the left end of
the QW. Similarly, the right reservoir with $\mu_2=eU_2$ 
injects a 
bare density 
\begin{equation}\label{bc2}
\rho_L^0(L/2)=eU_2/2\pi\hbar v_F
\end{equation}
of left-movers into the right end. 
These bare injected densities 
cannot depend on the intrinsic properties of the QW. In particular, 
they must be independent of the LL parameter $g$
and of the backscattering strength $\lambda$.
With the density of states $1/\pi \hbar v_F$,
and noting that a factor $1/2$ arises because only the left- or
right-moving density is injected, 
Eqs.~(\ref{bc1}) and (\ref{bc2}) readily follow.
The outgoing particle densities are not fixed by the
reservoirs. Outgoing particles
are assumed to enter ideal
reservoirs without reflection at the interface
between QW and reservoir.

According to Eqs.~(\ref{g2}) and (\ref{rmin}),
we can express the bare injected densities in 
terms of the true right- and left-moving densities,
\begin{eqnarray} \label{rel1}
\rho_R^0(x) &=& \frac{g^{-2}+1}{2} \rho_R(x) + 
\frac{g^{-2}-1}{2} \rho_L(x) \;, \\     \label{rel2}
\rho_L^0(x) &=& \frac{g^{-2}-1}{2} \rho_R(x) + 
\frac{g^{-2}+1}{2} \rho_L(x) \;.
\end{eqnarray}
From these relations and 
Eqs.~(\ref{bc1}) and (\ref{bc2}),
it is immediately clear that the external
reservoirs can be completely described in terms
of boundary conditions for the asymptotic true right- and 
left-moving densities $\rho_{R/L}$ in the QW. 
These boundary conditions should be imposed for the 
(ground-state or thermal) expectation value
of the densities.  As a short-hand notation, however,
the appropriate $\langle\cdots\rangle$ brackets are
mostly omitted in the sequel.

Employing Eq.~(\ref{bos}),
the densities $\rho_{R}$ and $\rho_L$ can now 
be expressed in terms of the displacement field $\theta(x,t)$,
\begin{eqnarray}
\rho_R+\rho_L&=& \frac{1}{\sqrt{\pi}} \partial_x\theta \;,\\
\rho_R-\rho_L&=& \frac{1}{\sqrt{\pi} v_F}\partial_t\theta\;.
\end{eqnarray}
Thereby we arrive at 
{\em radiative boundary conditions} for the displacement
field,
\begin{eqnarray} \label{bbc1}
\left(
\frac{1}{g^2} \partial_x + \frac{1}{v_F}\partial_t\right)
\langle \theta(x=-L/2,t) \rangle &=& \frac{eU_1}{\sqrt{\pi} \hbar v_F} \;,\\
\label{bbc2}\left(
\frac{1}{g^2} \partial_x - \frac{1}{v_F}\partial_t\right)
\langle \theta(x=L/2,t) \rangle &=& \frac{eU_2}{\sqrt{\pi} \hbar v_F}\;,
\end{eqnarray}
which have to be fulfilled
at all times $t$ in the stationary non-equilibrium state.
They hold provided ideal reservoirs
are {\em adiabatically} connected to the QW and
one is in the {\em low-energy regime}, 
where both the applied voltage $U=U_1-U_2$ 
and the temperature are very small compared
to the bandwidth. The latter is of the order
of the Fermi energy $E_F\approx \hbar v_F k_F$.
The consequences of the boundary conditions
(\ref{bbc1}) and (\ref{bbc2}) are investigated
in the next two sections. In the remainder of this
section, we focus on the two limiting cases
of perfect transmission and  perfect reflection.

Starting with the 
{\em clean case}, $\lambda=0$,
we first observe that all densities are $x$-independent
along the QW.  From Eqs.~(\ref{g2}) and 
(\ref{rmin}), the true right- and left-moving densities
are given by
\begin{eqnarray}
\rho_R &=& \frac12 (\rho_R^0-\rho_L^0) + \frac{g^2}{2} (
\rho_R^0+\rho_L^0) \;, \\
\rho_L &=& \frac12 (\rho_L^0-\rho_R^0) + \frac{g^2}{2} (
\rho_R^0+\rho_L^0) \;.
\end{eqnarray}
Even if no left-movers are injected $(U_2=0)$, the
shift of the band bottom due to the charging of the
gate-QW capacitor will induce a change in the 
density $\rho_L$ of left-movers.
These relations directly imply from Eq.~(\ref{ii}) 
the current
\begin{equation}
I=(e^2/h)\, U \;,
\end{equation}
which is the perfect conductance quantization
observed experimentally.\cite{tarucha} There is no
renormalization of the d.c.~conductance of a clean QW by
the electron-electron interaction.\cite{schulz,kubo,froh,prl}

The excess density $\rho=\rho_R+\rho_L$ charging
the gate-QW capacitor   is given by
\begin{equation}             \label{excess}
\rho= \frac{g^2 e(U_1+U_2)}{2\pi \hbar v_F} \;,
\end{equation}
and the electrostatic potential drop between
the QW and the backgate is then found from Eq.~(\ref{poisson}),
\begin{equation}
\varphi = (1-g^2) \frac{U_1+U_2}{2} \;.
\end{equation}
The rather incomplete screening in one dimension\cite{egger2} 
implies that only a fraction $(1-g^2)$ of the average potential
shift $(U_1+U_2)/2$ is compensated by the backgate,
leaving a fraction $g^2$ of the bare density as true charge density.
For a long-ranged $1/r$
interaction, one has $g\to 0$ in the long-wavelength
limit, and perfect electroneutrality ($\rho=0$) 
is recovered. In that case, the electrostatic potential
follows the chemical potential, $e\varphi= \mu$.
On the other hand, for the noninteracting case $g=1$,
the electrostatic potential vanishes, and the 
density is fully given by the injected density. We note that
for any $g$, there is no electric field acting along
the QW since the electrostatic potential is 
constant. The current flowing through the 
QW is of purely chemical origin.

Next we turn to the case of {\em perfect  reflection}, $\lambda\to
\infty$. Since no current can flow, we have $\rho_R(x)=\rho_L(x)$,
and Eqs.~(\ref{g2}) and (\ref{rmin}) then yield
\begin{eqnarray*}
\rho(x<0) &=& g^2 e U_1/ \pi \hbar v_F \;,\\
\rho(x>0) &=& g^2 e U_2/ \pi \hbar v_F \;.
\end{eqnarray*}
The density drop across the insulating barrier
is then given by
\begin{equation}   \label{drop}
\Delta \rho = \rho(x<0)-\rho(x>0) = g^2 eU/\pi\hbar v_F\;.
\end{equation}
From Eq.~(\ref{poisson}) we find the electrostatic
voltage drop across the barrier,
\begin{equation} \label{potdrop}
\Delta \varphi = (1-g^2) U \;,
\end{equation}
which is the applied two-terminal
voltage reduced  by the characteristic
underscreening factor $(1-g^2)$.
Note that the
potential drop between the QW
and the gate is $(1-g^2)U_1$ for $x<0$, and
$(1-g^2)U_2$ for $x>0$, respectively. 
This yields again Eq.~(\ref{potdrop}).
Of course, Eq.~(\ref{drop}) can be decomposed
into a chemical potential part and an electrostatic
part,
\begin{equation}
\Delta \rho = \frac{\Delta\mu-e\Delta\varphi}{\pi \hbar v_F}  \;,
 \end{equation}
where $\Delta \mu=\mu_1-\mu_2=eU$.
Electroneutrality is recovered only
for $g=0$, with $\Delta \varphi=U$. Finally, for $g=1$, there
is no electrostatic potential drop across the barrier.

\section{General effects of the voltage sources}
\label{sec:gen}

Next we discuss  general consequences of the
applied voltage $U=U_1-U_2$ for the system depicted
in Fig.~\ref{fig1}. Extending the reasoning of Ref.\onlinecite{prl}
to the real-time case, we
introduce a new field $q=\sqrt{4\pi}\theta(0)$
by means of a Lagrange multiplier field $\eta$. 
This has the advantage of rendering the $\theta(x)$ 
degree of freedom in a Gaussian form, and the
nonlinearity due to $H_{\rm imp}$ affects only $q$.  
We shall employ a path-integral representation in
the following.

Since it is convenient to integrate out the $\theta$ field,
all fields have to be defined on the Keldysh contour
${\cal C}$ extending from time $z=-\infty$ to $z=\infty$
(forward path) and back from $z=\infty$ to
$z=-\infty$ (backward path). For instance,
the field $q(z)$ consists of a forward path $q_f(t)$
and a backward path $q_b(t)$, where the time variable
$t$ now runs from $-\infty$ to $\infty$.
The action then reads
\begin{equation}\label{action1}
S= \int_{\cal C} dz\, L[\theta(z),q(z),\eta(z)]  \;,
\end{equation}
with the Lagrange function 
\begin{eqnarray}
L &=& \frac{\hbar v}{2g} \int dx \left[ \frac{1}{v^2} (\partial_z \theta)^2
- (\partial_x\theta)^2 \right] \\ \nonumber &-& \lambda \cos q(z) 
- \eta(z) [ q(z)- \sqrt{4\pi}\theta(0,z) ]   \;.
\end{eqnarray}
 The $\theta(x)$ field can now be
eliminated by Gaussian integration subject to the
radiative boundary conditions (\ref{bbc1})
and (\ref{bbc2}). This is achieved
by solving the Euler-Lagrange equation 
\begin{equation}  \label{euler}
(\frac{1}{v^2} \partial_z^2 - \partial_x^2)\theta(x,z) = 
\sqrt{4\pi}  \eta(z) \delta(x)/\hbar v_F  \;.
\end{equation}
The solution to this equation 
can always be decomposed into a particular solution
$\theta_p$ subject to the boundary conditions
plus the homogeneous solution obtained for $U_1=U_2=0$.
The latter is in fact well-known, see Ref.\onlinecite{kane}.
A particular solution obeying both Eq.~(\ref{euler}) 
and the boundary conditions 
(\ref{bbc1}) and (\ref{bbc2}) is
\begin{equation}\label{part}
\theta_p(x,z) = \frac{g^2 e[(U_1+U_2)x-V|x|]}{\sqrt{4\pi} \hbar v_F} 
    + \frac{e(U-V)z}{\sqrt{4\pi} \hbar}    \;,
\end{equation}
for both the forward and the backward path.
The quantity $V$  appears as the zero mode of
the Lagrange multiplier field $\eta(z)$. 
The physical meaning
of $V$ is the {\em four-terminal voltage} as becomes
 clear from the following discussion.

Since the expectation value of the
density operator $\rho(x)$ at $|x|\gg a$ 
 is determined by the particular
solution alone,\cite{prl} we obtain from 
 $\rho(x)=\partial_x\theta/\sqrt{\pi}$ the result
\begin{equation} \label{slow}
\langle\rho(x)\rangle= \frac{g^2 e(U_1+U_2)}{2\pi \hbar v_F}
- \frac{g^2 e V}{2\pi \hbar v_F} \,{\rm sgn}\, x \;.
\end{equation}
The first term is just Eq.~(\ref{excess}) 
describing the change in the overall 
charge density. It can be trivially  gauged to zero by choosing
$U_1=-U_2=U/2$.
The second term is more interesting. It gives the asymmetric charge
density in the presence of
an applied voltage. The density drop across the
barrier is thus
\begin{equation} \label{gdens}
\Delta \rho = g^2 eV/\pi \hbar v_F \;,
\end{equation}
such that there is an associated drop in the effective
chemical potential of size $\Delta\mu=g^2 eV$.
Equation (\ref{poisson}) then yields the
electrostatic potential drop at $x=0$,
\begin{equation} \label{vdens}
\Delta \varphi = (1- g^2) V \;.
\end{equation}
In a measurement of the four-terminal voltage,\cite{buttiker} the observed 
voltage drop is $\Delta\mu/e + \Delta\varphi$,
which is just $V$. Therefore $V$ is 
indeed the four-terminal voltage.
Since $V$ is introduced via
the Lagrange multiplier field $\eta$, it
is in general a fluctuating quantity. 
 
The ensuing steps are rather straightforward.
Since the technical details\cite{prl,grabert} are of no interest here, 
we will only sketch the analysis.
Solving Eq.~(\ref{euler}) for the 
homogeneous solution $\theta_h$ and
inserting $\theta=\theta_p+\theta_h$
back into $S$, one is left with a Gaussian
average over the Lagrange multiplier
field [except of the zero mode $V$, over which
we average separately].
Carrying out this Gaussian integration, we
obtain the effective action for averaging the local degree 
of freedom $q(z)$ and the four-terminal voltage $V$, 
\begin{eqnarray} \label{seff}
S_{\rm eff} &=&  i \Phi[q(z)] - \lambda\int_{\cal C} dz 
\,\cos [q(z)+e(U-V)z/\hbar]  \\ \nonumber
            &-& (eV/2\pi) \int_{\cal C} dz \, q(z) \;. 
\end{eqnarray}
The effects of the external voltage sources are contained in the
second and the third term. 
The first term can be written as
\begin{equation}\label{infl}
\Phi = \int_{\cal C} dz \int_{z>z'} dz'\, q(z) L(z-z') q(z') 
+ \frac{i A}{2} \int_{\cal C} dz\, q^2(z) \;,
\end{equation}
where $L(z)$ has the same form as 
the heat bath kernel in dissipative
quantum mechanics,\cite{uli}
\begin{equation}\label{lz}
L(z) =  \frac{\hbar}{\pi} \int_0^\infty d\omega \, J(\omega) \frac{
\cosh[\omega(-iz+\hbar\beta/2)]}{\sinh[\omega\hbar\beta/2]} 
\end{equation}
with $\beta=1/k_B T$.
The spectral density $J(\omega)$ is of Ohmic form,
\begin{equation}
J(\omega) = \frac{\omega}{2\pi g} \exp[-\hbar\omega/E_F] \;,
\end{equation}
where an exponential bandwidth cutoff has been chosen.
Finally, the quantity $A$ in Eq.~(\ref{infl}) is given by
\begin{equation}
A = \frac{2\hbar}{\pi} \int_0^\infty d\omega \, J(\omega)/\omega \;.
\end{equation}
The dissipation acting on $q(z)$ effectively comes from 
the eliminated degrees of freedom away from the
scatterer.\cite{kane} 

The effects of the applied voltage can now
be read off from Eq.~(\ref{seff}). 
The last term in $S_{\rm eff}$ is 
a {\em voltage drop} contribution 
obtained by making the assumption that
there is a local voltage drop $V$ 
at the impurity.  Under this assumption,
one can include the coupling to the voltage
sources by adding the term
\begin{equation} \label{voldrop}
\widetilde{H} = eV \theta(0)/\sqrt{\pi} 
\end{equation}
to the Hamiltonian.  Notably, it is 
in general not the externally applied
two-terminal voltage but the fluctuating
four-terminal voltage which determines
this part.  The second effect is a 
Josephson-like time dependence in the
argument of the second term in Eq.~(\ref{seff}).
Most importantly, because of this term
one cannot describe all effects 
of the applied voltage by simply adding
terms like Eq.~(\ref{voldrop}) to the Hamiltonian. 
In general, one has to solve the problem
under the radiative boundary conditions
(\ref{bbc1}) and (\ref{bbc2}).

Let us now briefly discuss the four-terminal voltage 
$V$. In the clean case, $\lambda=0$, the
field $q$ describes a massless particle
such that $V=0$ results from the associated
infrared divergence.  This is of 
course in accordance with Eq.~(\ref{gdens}),
since there is no density drop if there
is no barrier.  In the limit of perfect
reflection, $\lambda\to \infty$, the 
four-terminal voltage is $V=U$, as enforced by the
rapidly oscillating  impurity contribution in Eq.~(\ref{seff}).
This value can also be obtained by comparing Eqs.~(\ref{drop})
and (\ref{gdens}). As a function of $\lambda$,
the four-terminal voltage thus exhibits a crossover from
$V=0$ at $\lambda=0$ to $V=U$ for $\lambda\to\infty$.
Contrary to the Fermi liquid case, this crossover now
sensitively depends on the energy scales $k_B T$ and
$eU$ under consideration, see Sec.~\ref{sec:g12}.

The effective action (\ref{seff}) may 
serve as starting point for further
calculations, e.g., of the current-voltage
characteristics. We shall not pursue this
approach here but instead present an exact solution
for the special interaction 
strength $g=1/2$.

\section{Exact solution}
\label{sec:g12}

In this section we present  the exact solution
of the transport problem depicted in Fig.~\ref{fig1}
for the special LL parameter $g=1/2$.  This value
has been discussed previously,\cite{bcft,kane,g12,chamon,matveev}
essentially by assuming a local voltage drop term, i.e., by using
the effective action (\ref{seff}) under the assumption
$V=U$. However,  this assumption is only justified for  
a strong scatterer or at extremely low energy scales,
and one cannot recover the perfect conductance
$G=e^2/h$ of a clean QW using that approach.
Our exact solution for arbitrary
transmission reported below does not make the voltage drop assumption 
but instead uses the boundary conditions (\ref{bc1})
and (\ref{bc2}) to describe the coupling to the
reservoirs. Thereby the full crossover between the 
perfect conductance quantization and the asymptotic low-energy
localization due to the impurity is obtained.

To start, we introduce the chiral boson fields
\begin{eqnarray}
\varphi_R(x) &=& \sqrt{\pi} \left[ \frac{1}{\sqrt{g}} \theta(x)
+ \sqrt{g} \phi(x) \right]\;, \\
\varphi_L(x) &=& \sqrt{\pi} \left[ -\frac{1}{\sqrt{g}} \theta(x)
+ \sqrt{g} \phi(x) \right] \;.
\end{eqnarray}
According to Eq.~(\ref{algebra}), they obey the algebra ($p=R,L=\pm$) 
\begin{equation}
[ \varphi_p(x), \varphi_{p'}(x')]_- = -i\pi p \delta_{pp'}
\,{\rm sgn}\, (x-x') \;.
\end{equation}
The right- and left-moving densities in the QW are
\begin{equation}\label{rld}
\rho_{R,L}(x) = \pm \frac{\sqrt{g}}{2\pi} \partial_x \varphi_{R,L}(x) \;,
\end{equation}
and the Hamiltonian $H=H_{\rm LL}+H_{\rm imp}$ reads 
\begin{eqnarray}\label{hamg}
H&=&\frac{\hbar v}{8\pi}\int dx\left\{ (\partial_x\varphi_R)^2
+  (\partial_x\varphi_L)^2 \right\} \\ \nonumber &+&
\lambda \cos\{\sqrt{g}[\varphi_R(0)-\varphi_L(0)]\}   \;.
\end{eqnarray}

Next we incorporate the applied voltage sources
according to the boundary conditions (\ref{bc1})
and (\ref{bc2}). Using the relations (\ref{rel1})
and (\ref{rel2}), they lead to the conditions
\begin{eqnarray}\label{bcg}
(g^{-2}+1)\rho_R(-L/2)   
+ (g^{-2}-1)\rho_L(-L/2)  &=& \frac{eU_1}{\pi\hbar v_F} \;,\\
\nonumber (g^{-2}-1)\rho_R(L/2) 
+ (g^{-2}+1)\rho_L(L/2) &=& \frac{eU_2}{\pi \hbar v_F} \;.
\end{eqnarray}
It is then of advantage to switch to new chiral right-moving
fields defined by
\begin{equation}\label{newch}
\phi_{p=\pm}(x) = \frac{1}{\sqrt{2}} \left[ \varphi_R(x)
\mp \varphi_L(-x) \right] \;,
\end{equation}
subject to the algebra 
\begin{equation}
[ \phi_p(x), \phi_{p'}(x')]_- = -i\pi \delta_{pp'}
\,{\rm sgn}\, (x-x') \;.
\end{equation}
They define the densities
\begin{eqnarray}\label{newdens}
\widetilde{\rho}_{\pm}(x) &=& \frac{1}{2\pi} \partial_x \phi_{\pm}(x) \\
\nonumber     &=& \frac{1}{\sqrt{2g}} \{ \rho_R(x) \mp \rho_L(-x) \} \;.
\end{eqnarray}
Thereby the boundary conditions (\ref{bcg}) become
conditions for the new chiral densities (\ref{newdens}). 
Specializing on $g=1/2$, 
and taking the sum and difference of the emerging equations, 
we obtain 
\begin{eqnarray}\label{sum}
5\widetilde{\rho}_-(-L/2)+ 
 3\widetilde{\rho}_-(L/2)  
&=&e(U_1+U_2)/\pi \hbar v_F \;,
\\ \label{diff}
5\widetilde{\rho}_+(-L/2)- 
 3\widetilde{\rho}_+(L/2)  
&=&eU/\pi \hbar v_F \;.
\end{eqnarray}
The Hamiltonian (\ref{hamg}) 
expressed in terms of the new chiral
fields for $g=1/2$ is
\begin{equation}\label{hamg2}
H=\frac{\hbar v}{8\pi}\int dx\left\{ (\partial_x\phi_+)^2
+  (\partial_x\phi_-)^2 \right\} +
\lambda \cos[\phi_+(0)]   \;.
\end{equation}
It is now apparent that the $\phi_\pm$ fields are completely
decoupled. The impurity term in the Hamiltonian (\ref{hamg2})
couples only to $\phi_+$, and the applied voltage $U$ 
also  leads to a boundary condition only in the $(+)$ sector,
see Eq.~(\ref{diff}).
The $\phi_-$ field is associated with the shift in 
the total density arising for $U_1\neq -U_2$.
Since there is no backscattering in the $(-)$ sector,
the density $\widetilde{\rho}_-(x)$ stays constant along the QW,
and, according to Eq.~(\ref{sum}), we again obtain the
excess density (\ref{excess}) injected by the 
reservoirs. This shift in the 
overall density does not lead to interesting physical effects.
Putting $U_1=-U_2=U/2$, we only keep the 
$\phi_+$ field in what follows.

By means of refermionization,\cite{kane,g12,chamon,matveev}
we can then obtain an exact solution.
For that purpose, we first introduce  new fermion operators
\begin{equation}\label{ref}
\widetilde{\psi}(x) = (2\pi a)^{-1/2} \exp[i\phi_+(x)] \;.
\end{equation}
Following Matveev,\cite{matveev} it is convenient
to  switch in a second step to the fermion operators $\psi$
defined by 
\begin{equation}
\widetilde{\psi}(x) = (c^{}+c^\dagger) \psi(x) \;,
\end{equation}
where $c$ is an auxiliary fermion.
Expressed in terms of these fermion operators, the 
$(+)$ sector of the Hamiltonian (\ref{hamg2}) reads
\begin{eqnarray}\label{hamg3}
H &=& -i \hbar v\int dx \, \psi^\dagger(x) \partial_x \psi^{}(x)
\\ \nonumber &+& ( \hbar v\lambda_B/2)^{1/2} (c^{}+c^\dagger) \left[
\psi^{}(0)-\psi^\dagger(0) \right] \;,
\end{eqnarray}
with the effective impurity strength 
\begin{equation} \label{effimp}
\lambda_B = \pi a \lambda^2/\hbar v\;.
\end{equation}
Remarkably, in the refermionized version (\ref{hamg3}) 
the Hamiltonian attains a very simple form,  which can
be diagonalized by, e.g., the equation-of-motion
method.\cite{chamon}  
Switching to Fourier space, 
\begin{equation}                   \label{fourier}
\psi(x,t) = \frac{1}{L} \sum_k \, \exp[ik(vt-x)] 
\times \left\{ \begin{array}{l} 
a_k \quad   (x<0) \\ b_k \quad (x>0) \end{array} \right. \;,
\end{equation}
where $k$ runs over integer multiples of $2\pi/L$ and
$a_k, b_k$ denote fermion operators,
the equations of motion dictate\cite{chamon,foot}
\begin{equation}  \label{bk}
b^{}_k = \frac12 \left( 1+e^{i\alpha_k} \right) a^{}_k 
  + \frac12 \left( 1-e^{-i\alpha_k} \right) a^\dagger_{-k}\;,
\end{equation}
where the scattering phase  shift $\alpha_k$ is defined
by
\begin{equation}   \label{phase}
e^{i\alpha_k} = e^{-i\alpha_{-k}} 
= \frac{i\hbar vk-\lambda_B}{i \hbar vk+\lambda_B}          \;.
\end{equation}

So far the analysis has closely followed previous work,
see, e.g., Ref.\onlinecite{chamon}.  Now we have to take
into account the boundary condition (\ref{diff})
in order to incorporate the applied voltage $U$.
First we note that the density operator $\widetilde{\rho}_+(x)$
defined in Eq.~(\ref{newdens}) can equivalently be expressed
in terms of the new fermion operator $\psi(x)$,
\begin{equation}
\widetilde{\rho}_+(x) = \psi^\dagger(x)\psi^{}(x) \;.
\end{equation}
Employing Eq.~(\ref{fourier}), the  
boundary condition (\ref{diff}) then leads to 
\begin{equation} \label{bcfinal}
\frac{1}{L}\sum_k \left\{ 5 \langle a_k^\dagger a_k^{} 
\rangle^\prime  - 3 \langle b_k^\dagger b_k^{} \rangle^\prime
\right\} = eU/\pi \hbar v_F \;.
\end{equation}
The brackets indicate a stationary nonequilibrium 
average, and the prime stands for
normal-ordering with respect to
the $U=0$ equilibrium state.
Since the $a_k$ correspond to free fermions, they must
obey the Fermi distribution function,
\begin{equation}
    \langle a_k^\dagger a_k^{} \rangle \equiv n_k(k^*) = 
    [1+\exp\{\hbar \beta v(k-k^*)\}]^{-1} \;,
\end{equation}
where $k^*$ has to be determined
self-consistently. Using  Eq.~(\ref{bk}), we obtain
\begin{equation}
\langle b_k^\dagger b_k^{} \rangle =
\frac12 (1+\cos\alpha_k) n_k(k^*) +
\frac12 (1-\cos\alpha_k) n_k(-k^*)     \;,
\end{equation}
whence Eq.~(\ref{bcfinal}) with the scattering phase shift
(\ref{phase}) yields
\begin{equation} \label{fin}
k^* + \frac{6\pi}{L} \sum_k \, [1+(\hbar vk/\lambda_B)^2]^{-1}
\, \{ n_k(k^*)-n_k(0) \} = \frac{eU}{\hbar v_F } \;.
\end{equation}
In the remainder, we focus on the case  
of a very long QW, $L\to \infty$,
such that sums can be converted into 
integrals, $(2\pi/L)\sum_k \to \int dk$.  
Carrying out the resulting integration, the
condition (\ref{fin}) reads 
\begin{equation}\label{cond2}
k^* + \frac{3\lambda_B}{\hbar v}\;{\rm Im}\,
\psi\left(\frac12 + \frac{\lambda_B+i
\hbar vk^*}{2\pi k_B T}
\right) = eU/\hbar v_F  \;,
\end{equation}
where $\psi(z)$ is the digamma function.
For $\lambda_B=0$, this gives $k^*=eU/\hbar v_F$,
while in the opposite limit of a strong scatterer,
$\lambda_B\to \infty$, we obtain $k^*=eU/4\hbar v_F$.
These two extreme values hold in fact for any value
of the temperature $T$ or the length $L$. 
The crossover as a function of $\lambda_B$
between these two limits strongly depends
on the energy scales $k_B T$ and $eU$. 
Clearly, for $\lambda_B\gg  eU$, we could effectively 
use the strong-coupling value $k^*=eU/4\hbar v_F$.
This amounts to making the above-mentioned
voltage drop assumption.
In the general case, one first has to solve for $k^*$  
according to Eq.~(\ref{cond2}) before further calculations.

Let us now study the connection to the {\em four-terminal
voltage} $V$ discussed in the previous section.
It can be obtained from the density drop 
$\Delta\rho=\rho(x<0)-\rho(x>0)$
at $x=0$. Using Eq.~(\ref{newdens}), 
we have
\begin{equation}
\Delta\rho= \widetilde{\rho}_+(x<0)
- \widetilde{\rho}_+(x>0) = \Delta\widetilde{\rho}_+ \;,
\end{equation}
which yields
\begin{eqnarray} \nonumber
\Delta\rho &=& \int \frac{dk}{2\pi} \langle [a_k^\dagger a^{}_k
- b_k^\dagger b_k^{}] \rangle^\prime    \\
\nonumber 
&=& \int \frac{dk}{2\pi} \,(1-\cos\alpha_k) \{ n_k(k^*)-n_k(0) \} \\
&=& \frac{\lambda_B}{\pi \hbar v}
 \;{\rm Im}\,
\psi\left(\frac12 + \frac{\lambda_B+i\hbar
vk^*}{2\pi k_B T} \right) \;.
\end{eqnarray}
Comparing with the general result (\ref{gdens}), the 
 four-terminal voltage $V$ follows,
\begin{equation}\label{ftv2}
eV = 2\lambda_B 
 \;{\rm Im}\,
\psi\left(\frac12 + \frac{\lambda_B+i
\hbar vk^*}{2\pi k_B T} \right) \;.
\end{equation}
The generalization to finite length $L$ is straightforward.
From our exact solution, one can in principle 
also compute the {\em fluctuations} of the four-terminal voltage. 
When comparing with experiments, however, one may
have to include the strong Friedel oscillation
contribution.\cite{buttiker}

In the limit of a clean wire, from Eq.~(\ref{ftv2})
we find  $V=0$, in accordance
with the general result for arbitrary $g$. In the 
opposite case, $\lambda_B\to \infty$, we obtain $V=U$ 
from Eqs.~(\ref{cond2}) and (\ref{ftv2}), again in 
accordance with the general result. The 
connection between $k^*$ and $V$ can now be read off,
\begin{equation} \label{cond3}
k^* = e[U-3V/4]/\hbar v_F \;.
\end{equation}
We stress that this relation holds for any $T$ and $L$.
Inserting Eq.~(\ref{cond3}) into Eq.~(\ref{ftv2}),
we can eliminate $k^*$ and obtain a 
self-consistent equation for the four-terminal
voltage,
\begin{equation}\label{ftv3}
eV/2\lambda_B =  
\; {\rm Im}\,
\psi\left(\frac12 + \frac{\lambda_B+2ieU-3ieV/2}{2\pi k_B T} \right) \;.
\end{equation}
At zero temperature, this becomes
\begin{equation}\label{t0}
eV/2\lambda_B = \tan^{-1}\{[2eU-3eV/2]/\lambda_B\} \;.
\end{equation}
The relation (\ref{ftv3}) explicitly exhibits {\em scaling} with the
effective impurity strength (\ref{effimp}) acting as
the energy scale, i.e., the energies $k_B T$, $eU$, and
$eV$ can be turned into dimensionless quantities
by measuring them in units of $\lambda_B$. 
Therefore the boundary conditions preserve the
important scaling property.  For small $\lambda_B$,
the four-terminal voltage $V$ vanishes, and
by increasing $\lambda_B$, a crossover to the
strong-coupling value $V=U$ is observed.

\begin{figure}
\epsfysize=7cm
\epsffile{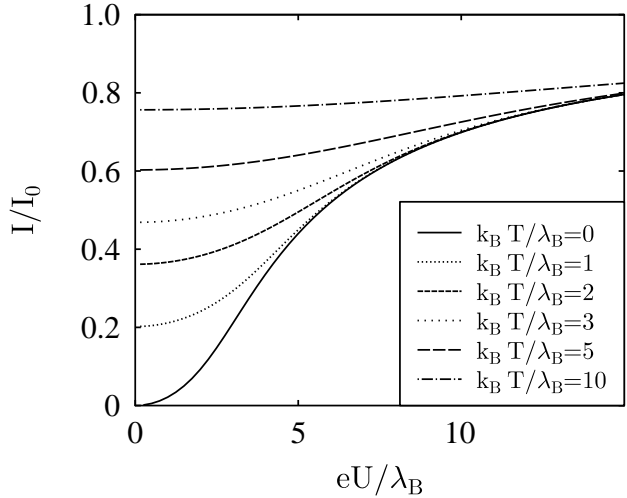}
\caption[]{\label{fig2}
Current-voltage characteristics for
several temperatures $T$.  The current has been
normalized to $I_0=(e^2/h)U$.  Note the very slow
approach towards $I=I_0$ as temperature is raised.} 
\end{figure}

Finally, we come to the {\em current-voltage characteristics}.
The current flowing through the QW is computed 
from Eq.~(\ref{ii}),
\begin{eqnarray*}
I &=& ev_F \langle \widetilde{\rho}_+(0) \rangle \\
&=& \frac{ev_F}{4} \int \frac{dk}{2\pi}
\langle (a_k^\dagger+b_k^\dagger)(a_k^{}+b_k^{}) \rangle^\prime \;.
\end{eqnarray*}
Straightforward algebra yields   the general result
\begin{equation}\label{iv}
I(U) = \frac{e^2}{h} (U-V) \;,
\end{equation}
with the four-terminal voltage $V=V(U,T,\lambda_B)$ 
self-consistently given in Eq.~(\ref{ftv3}). 
Therefore the knowledge of the four-terminal voltage
is sufficient to obtain the full nonlinear
current-voltage characteristics.
In the limit of a clean QW, $V=0$, and we indeed obtain
the conductance quantum $G=e^2/h$. 
In the limit of very small applied voltage, $eU\ll \lambda_B$,
and at zero temperature, the voltage drop
assumption is correct, and the previous results\cite{bcft,kane,g12,chamon}
are recovered. 

  The exact current-voltage characteristics
is plotted in Fig.~\ref{fig2} for various temperatures.
Clearly, one has a perfect zero-bias anomaly at $T=0$,
with the conductance vanishing $\sim U^2$ as predicted
by Kane and Fisher.\cite{kane}  Notably, 
Eq.~(\ref{iv}) gives the full crossover 
behavior up to the perfect conductance $G=e^2/h$ of a clean QW.

\section{Conclusions}
\label{sec:conc}

In this paper, the inclusion of external voltage
sources to a one-dimensional single-channel
quantum wire with arbitrary 
transmission has been discussed.  This system
is a prototypical example for a Luttinger liquid.
By deriving radiative boundary conditions, we have 
demonstrated that  the Landauer approach to
mesoscopic transport can be extended
to the case of strongly correlated 
systems.  The exact solution of the
transport problem at the special
value $g=1/2$ reveals that both the 
previous ``voltage drop'' results (which
hold at sufficiently low voltage and temperature)
and the perfect conductance quantization 
in a clean system can be recovered within a
unified approach.

An obvious and interesting generalization
 concerns the a.c.~case.
Considering a situation where $U_1= U\cos(\omega t)$
and $U_2=0$, the boundary condition at the left
end of the wire would read
\begin{equation}
\rho_R^0 = \frac{eU_1 \cos(\omega t - \omega x/v)}{2\pi \hbar v_F} \;.
\end{equation}
The consequences of time-dependent boundary conditions
have not been studied so far except in
the clean case.\cite{blanter}
Our boundary condition approach also allows for
a consideration of more complicated geometries. 
 For instance, the 
problem of crossed Luttinger liquids allows
for an elegant solution by 
employing this approach.\cite{komnik}

\acknowledgements 

The authors would like to thank A. Komnik for a critical reading
of the manuscript.  Support has been provided by the Deutsche
Forschungsgemeinschaft (Bonn).


\begin{references}
\bibitem{forfoot} 
There is also a forward scattering term caused by the
impurity which can be removed by a unitary transformation
and does not affect the physics discussed here.
\bibitem{foot} 
Directly at $x=0$ there is a discontinuity in the fermion 
operator $\psi(x)$ due to a jump condition from the
equation of motion.
To ensure the proper anticommutation algebra,
the fermion operator at $x=0$ must be defined
as $\psi(0)=\frac12 [\psi(0^-)+\psi(0^+)]$.
\end{references}
\end{document}